# Where Does Dark Matter Become Important in Elliptical Galaxies

Hans-Walter Rix

*Steward Observatory, University of Arizona and Max-Planck-Institute für Astrophysik, Garching*

## 1. What are the Questions?

Even though the construction of self-consistent galaxy models, where all the mass is provided by the luminous matter is still viewed as the holy grail by many stellar dynamicists, there can be little doubt that elliptical galaxies are ultimately surrounded by dark matter halos which dominate the dynamics at large radii. While the radial mass distribution in numerous spiral galaxies has been mapped out in detail (see *e.g.* Persic *et al.* 1996 or Broeils and Courteau, this volume), many basic question regarding the halos of ellipticals remain unanswered:

- At what radius are luminous and dark matter comparably important, *i.e.* $M_\star(< R) \approx M_{dark}(< R)$?

- Is the rotation curve, $v_c \equiv \sqrt{R \partial \Phi / \partial R}$, "flat" across the transition from predominantly luminous to dark matter?

- Do spiral galaxies and elliptical galaxies of the same *stellar* mass reside in comparable halos?

These questions cannot be answered by simple analogy with spiral galaxies, because (1) the stellar bodies of ellipticals are two orders of magnitude more compact than those of spirals with the same stellar mass; (2) luminous ellipticals are the most massive single galaxies; (3) ellipticals formed in a different environment and with more violent relaxation.

We know much less about elliptical galaxies because there is no single kinematic tracer that is present in all objects and that covers a wide range of radii (as the neutral and ionized gas in spirals). For very massive ellipticals the X-ray gas (*e.g.* Forman, Jones and Tucker, 1985) provides a solid estimate of the mass enclosed within $\sim 10 R_{eff}$ and for very few objects (*e.g.* NGC 4278, Lees 1992; IC 2006, Franx *et al.* 1994) there is an HI ring for mass estimates. For a similarly small number of objects, gravitational lensing has provided very accurate estimates of the projected mass enclosed within $\sim R_{eff}$. However, for the majority of objects, especially for galaxies fainter than $L_\star$, we still need to rely on the stellar kinematics to probe the potential. For a number of reasons it has so far proven difficult, if not impossible, to probe the dark matter content of ellipticals by stellar dynamical means: Kinematic data were only available inside $R_{eff}$, and only the first two moments of the stellar line-of-sight velocity distribution (LVD), $v$ and $\sigma$, could be extracted from the data and were available for comparison with the models. Further, many models had to *assume* a model anisotropy, rather than have it be constrained by the data. As we will describe below significant progress in all these areas is underway, greatly increasing the use of stellar kinematics in constraining the dark matter.

## 2. Gravitational Lensing

The strength and weakness of gravitational lensing is that it probes one particular radius. The strength lies in the exquisite accuracy with which the mass enclosed within the Einstein ring radius can be measured (*e.g.* Kochanek 1991, Rix *et al.* 1992), typically a few percent. On the other hand, the only other information that can be extracted about the mass profile is its gradient at that same radius, $\partial M/\partial R$. Lensing can probe the mass distribution statistically (*e.g.* Maoz and Rix, 1992), or through detailed models of the few cases where an extended sources is lensed (*e.g.* MG1131+0456 and MG1654+13, Kochanek 1995, Chen *et al.* 1995). These studies find that the mass-to-light ratios at $R_{eff}$ are slightly higher than inferred from the central velocity dispersion and that the constant mass-to-light hypothesis is inconsistent with the data. Without a halo the predicted distribution of image separations for lensed QSOs are smaller than the observed distribution (Maoz and Rix, 1992). More clearly, the studies of MG1131+0456 and MG1654+13 show that $M(<R)$ increases linearly with $R$ ar $R_{eff}$. This leads to a deprojected density profile of $\rho_{mass} \propto r^{-2}$, at a radius where the stars fall off as $\rho_\star \propto r^{-3}$. This shows clearly that in at least these few objects the dark matter becomes important at an effective radius and seems to lead to a "flat" rotation curve.

## 3. Stellar Dynamics: Modeling the Stellar Velocity Distribution

Despite X-ray measurements, planetary nebulae (*e.g.* Ciardullo *et al.* 1993) and gravitational lensing, stellar dynamics remain a principal probe of the potential. Its main advantages lie in the ubiquity of a well sampled tracer and the large radial range (factor $\sim 100$) probed. The difficulties lie in solving simultaneously for *both* the potential in which the stars orbit *and* the statistical distribution of orbit properties. Previously available data either covered only radii inside $R_{eff}$, or only determined $v$ and $\sigma$. As a consequence, models based on a considerable range of potentials could match the data with a variety of orbital distributions. However, many of these models lead to significantly non-Gaussian velocity distributions, are "pay the price" at larger (usually unobserved radii). In the last years it has become possible to measure the full line-of-sight velocity distribution (LVD) of the stars from absorption line spectra (*e.g.* Rix and White, 1992), and to extend such LVD data well beyond $R_{eff}$ (Carollo *et al.* 1995). It often proves convenient to parameterize the shape of the LVD by Gauss Hermite moments, $h_i$, where, for example, $h_4 > 0$ indicates a line profile more peaked than a Gaussian.

### 3.1. Updating Schwarzschild's Method

With this new wealth of data now available, the need for a commensurate modeling technique arises. To this end we (HWR, T. de Zeeuw, R. van der Marel, M. Carollo and N. Cretton) decided to update Schwarzschild's (1979) method of building galaxies from the non-negative superposition of time-averaged orbits. Conceptually our method has few differences compared to earlier implementations (*e.g.* Richstone and Tremaine, 1984), but we have included a number of features (Rix *et al.* 1997) which make the method more powerful in practice:

- the model can match simultaneously the photometry and the full LVD.

- a proper seeing convolution is implemented.

- the modeling takes full account of the observational errors in the photometry and the kinematics; in this way we can calculate the relative likelihoods for any set of trial potentials, given the observational constraints.

- using the likelihood approach permits to determine the smoothest distribution function that is *equally consistent with the data*.

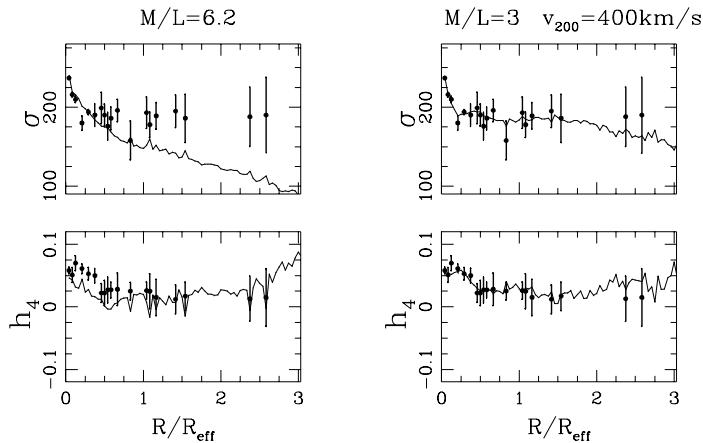

Figure 1. Dynamical Models for NGC 2434 without and with a dark halo. The data for the round and nearly non-rotating galaxy were taken from Carollo *et al.* (1995) and were modeled with the method described in §3.1. When the LVD is forced to match the LVD shape ($h_4$), the constant M/L models fail to fit the data. However, a model with a dark halo, yielding an effectively flat rotation curve fits the data well.

### 3.2. The Choice of Dark Matter Potentials

Combining this updated modeling with LVD data that extend to several $R_{eff}$ (from Carollo *et al.* 1995, and Rix 1997) has yielded some intriguing results: As is shown in Figure 1a, for NGC 2434, the constant M/L can be clearly rejected, once the line profile shape is constrained. Similar results were found for NGC 7619. Only for very few cases can the stellar dynamical modeling be combined with gas kinematics to actually map the rotation curve. In general, the non-local nature of the orbits forces us to provide a sequence of model potentials to be tested against the data. For spiral galaxies the traditional solution is to parameterize the halo by a non-singular isothermal sphere. This approach introduces two further parameters and is not physically motivated. As an alternative, we explore the halo profiles suggested by Navarro *et al.* (1995, NFW), based on their cosmological structure formation simulations. For any given cosmogony (*e.g.* SCDM) NFW find a one parameter sequence of halo profiles that can be labeled by their mass scale or circular velocity, $v_{200}$. These halo profiles are "cuspy", following $\rho \propto r^{-1}$ at small radii and $\rho \propto r^{-3}$ at large radii. The NFW simulations do not involve a dissipative mass components, and we have to account for the adjustment of the halo profiles due to the baryons concentrating at their center. We do this by assuming adiabatic contraction (Blumenthal *et al.* 1984), which is justified if the baryonic mass concentrated at the center on a timescale longer that the characteristic dynamical time ($\sim 10^8$ yrs).

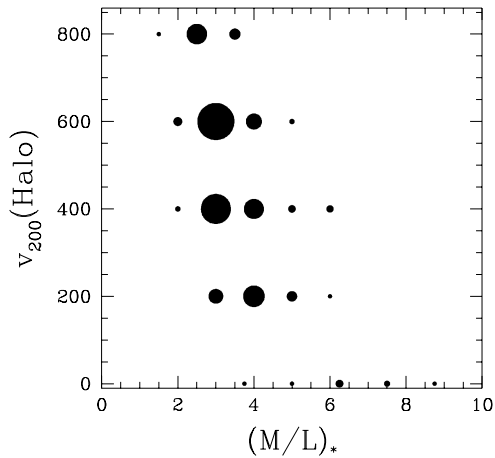

Figure 2. Relative likelihood of models for NGC 2434 in the $M/L_\star$ vs. $v_{200}$ plane. The covariance between stellar and halo mass is apparent. The best model is found trough interpolation of the grid $v_{200} = 420 \pm 90$km/s and $M/L_\star(V) = 3.5 \pm 0.5$.

### 3.3. Results for NGC 2434 and NGC 7619

With the above procedure we can create a consistent and well motivated set of total potentials, described by the stellar mass-to-light ration $M/L_\star$ and the halo velocity scale $v_{200}$. For a grid in $M/L_\star - v_{200}$ we calculate a set of orbits, find the best superposition to match the observed photometry and velocity distributions, and calculate $\chi^2$ for the best fit. The $\chi^2$ values in the $M/L_\star - v_{200}$ plane, converted to relative likelihoods are shown in Figure 2 for NGC 2434. The best fit for NGC 2434 is $M/L_\star(V) = 3.5$ and $v_{200} = 420$ is shown in Figure 1b. This figure clearly shows the covariance between stellar and dark mass, roughly preserving the total mass inside $\sim R_{eff}$. However, through the shape of the rotation curve the halo mass $v_{200}$ by itself is constrained: for NGC 2434 we find $v_{200} = 420 \pm 90$km/s, and for NGC 7619 a similar analysis yields $v_{200} = 450 \pm 120$km/s (We use a concentration parameter for the NFW halos of $\log c = 1$, corresponding to a standard CDM cosmogony). Figure 4 shows the stellar, dark matter and total circular velocity for the best fitting NGC 2434 model and illustrates a number of interesting points: (1) the circular velocity curve remains virtually constant from 0.2 to $3R_{eff}$, across the transition to dark matter domination; (2) half of the total mass inside the effective radius is dark; (3) for cuspy halo models the M/L changes continuously starting at the center.

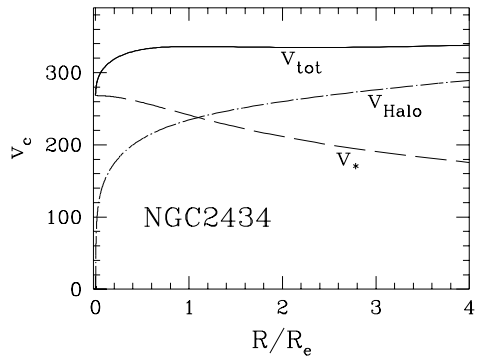

Figure 3. The "rotation curve", $R\partial\Phi/\partial R$, of the best fitting model for NGC 2434 (from Rix *et al.* 1997), using the star – halo models from §3.2. The total rotation curve is flat for $0.3 < R/R_{eff} \leq 4$, and the contributions from the stars and the dark matter are equal at the effective radius.

## 4. Dense Halos around Ellipticals?

The two galaxies analyzed show that combining the new data on stellar velocity distributions with appropriate modeling can provide significant new constraints on the mass distribution in ellipticals to $\geq 2R_{eff}$. The preliminary results on NGC 2434 and NGC 7619 and the lensing analyses (§2) appear to shape themselves into a coherent picture, providing preliminary answers to the questions posed initially: The total mass distribution appears have the profile $\rho_{tot} \propto r^{-2}$ near the effective radius, where $\rho_{lum} \propto r^{-3}$. At the same radius luminous and dark matter contribute equally to the potential. Because ellipticals are such dense systems ($R_{0.5} \sim 1.5$ kpc for a $0.5L_*$ galaxy, compared to the half light radius $R_{0.5} \sim 7$ kpc, for an $L_*$ spiral of comparable stellar mass), they must have very dense halos. Navarro (this volume) finds that spirals are best fit by lower density halos ($\log c \approx 0.5$, arising *e.g.* in a low density universe). To fit the the stellar kinematics of the two ellipticals with $\log c = 0.5$ would require halos of $v_{200} \sim 1000$ km/s. Whether this is in direct conflict with observations remains to be checked. Alternatively, the higher density of halos around ellipticals may result from their formation at higher redshift.

Clearly, detailed studies for a larger number of galaxies are required. But we should feel encouraged, because recent progress in gravitational lensing, stellar dynamics of ellipticals and cosmological halo formation simulations may interact to provide a powerful probe of cosmogonies and galaxy formation.

It is a pleasure to thank my collaborators for allowing me to present results in advance of publication. I am grateful for fruitful discussions with J. Navarro and D. Weinberg and I would like to thanks the organizers for the enjoyable and productive workshop.